%% file: main.tex
\documentclass[runningheads]{llncs}
\usepackage[T1]{fontenc}
\usepackage{graphicx}
\usepackage{hhline}
\usepackage{multirow}
\usepackage[ruled]{algorithm2e}
\usepackage{qtree}
\usepackage{listings}
\usepackage{xcolor}
\definecolor{darkgreen}{RGB}{13, 74, 33}
\definecolor{darkturquoise}{RGB}{28, 67, 76}
\definecolor{darkbrick}{RGB}{120, 0, 26}
\definecolor{darkyellow}{RGB}{166, 89, 0}
\definecolor{darkgrey}{RGB}{50, 50, 50}
\definecolor{green}{RGB}{77, 160, 97}
\definecolor{turquoise}{RGB}{51, 156, 156}
\definecolor{brick}{RGB}{232, 106, 88}
\definecolor{yellow}{RGB}{255, 190, 0}
\definecolor{grey}{RGB}{165, 165, 165}
\definecolor{lightgreen}{RGB}{199, 235, 186}
\definecolor{lightturquoise}{RGB}{178, 224, 224}
\definecolor{lightbrick}{RGB}{255, 204, 196}
\definecolor{lightyellow}{RGB}{255, 240, 176}
\definecolor{lightgrey}{RGB}{230, 230, 230}
\lstdefinelanguage{model}{
    morecomment=[l]{\#},
    keywords = {operation, constraint}
}
\lstdefinelanguage{asm}{
    morecomment=[l]{\#},
    keywords = {@C,@S,@R,@A,@W,@I}
}
\lstdefinelanguage{computation_asm}{
    morecomment=[l]{\#},
    keywords = {load, compare, branch, add, jump, nop}
}
\lstdefinelanguage{resource_asm}{
    morecomment=[l]{\#},
    keywords = {repeat, forever}
}

\begin{document}
\title{CIS: Composable Instruction Set\\for Data Streaming Applications}
%
%
\author{Yu Yang\inst{1}\orcidID{0000-0003-2396-3590} \and
Jordi Altayó González\inst{1}\orcidID{0000-0002-7693-6994} \and
Paul Delestrac\inst{1}\orcidID{0000-0002-7476-1422} \and
Ahmed Hemani\inst{1}\orcidID{0000-0003-0565-9376}}
\authorrunning{Y. Yang et al.}
%
\institute{KTH Royal Institute of Technology, SE-100 44 Stockholm, Sweden
\email{\{yuyang2,jordiag,pauldel,hemani\}@kth.se}}

\maketitle              

\input{contents/abstract}
\input{contents/sec1}
\input{contents/sec2}
\input{contents/sec3}
\input{contents/sec4}
\input{contents/sec5}

\input{contents/ack}
%
%
%
\bibliographystyle{splncs04}
\bibliography{contents/reference}

\end{document}

%% file: contents/abstract.tex
\begin{abstract}
The enhanced efficiency of hardware accelerators, including Single Instruction Multiple Data (SIMD) architectures and Coarse-Grained Reconfigurable Architectures (CGRAs), is driving significant advancements in Artificial Intelligence and Machine Learning (AI/ML) applications. These applications frequently involve data streaming operations comprised of numerous vector calculations inherently amenable to parallelization. However, despite considerable progress in hardware accelerator design, their potential remains constrained by conventional instruction set architectures (ISAs). Traditional ISAs, primarily designed for microprocessors and accelerators, emphasize computation while often neglecting instruction composability and inter-instruction cooperation. This limitation results in rigid ISAs that are difficult to extend and suffer from large control overhead in their hardware implementations. To address this, we present a novel composable instruction set (CIS) architecture, designed with both spatial and temporal composability, making it well-suited for data streaming applications. The proposed CIS utilizes a small instruction set, yet efficiently implements complex, multi-level loop structures essential for accelerating data streaming workloads. Furthermore, CIS adopts a resource-centric approach, facilitating straightforward extension through the integration of new hardware resources, enabling the creation of custom, heterogeneous computing platforms. Our results comparing performance between the proposed CIS and other state-of-the-art ISAs demonstrate that a CIS-based architecture significantly outperforms existing solutions, achieving near-optimal processing element (PE) utilization.
\end{abstract}

\keywords{ISA, data streaming application, heterogeneous accelerator, composable instruction}

%% file: contents/sec1.tex
\section{Introduction}\label{sec:1}
Streaming applications, dominated by data-parallel vector operations, have been widely used in AI and Machine Learning, signal processing, multimedia, and many other application domains. Accelerating these data streaming applications is essential. The key to achieving this goal relies on improving the execution of the static loops that dominate the data steaming application operations, by shifting away from traditional \emph{computation-centric instructions} and embracing \emph{resource-centric instructions}.  We show a motivational example to introduce the concept of resource-centric instructions and their benefits compared to traditional computation-centric instructions.

\begin{minipage}{\linewidth}
\begin{lstlisting}[language=c, caption=A static C program, label=list:c_program, xrightmargin=4.0ex]
for(i=0; i<64; i++) {
  A[i] = A[i] + 1;
}
\end{lstlisting}
\end{minipage}

Consider the simple C program in Listing~\ref{list:c_program}, where we perform an ``ADD-1'' operation 64 times. The conventional processor compiler will translate the C program into assembly code similar to Listing~\ref{list:computation_asm}. The instructions for traditional processors and the majority of configurable accelerators are computation-centric (i.e., each instruction mobilizes all the necessary resources to implement the complete instruction functionality). For example, the key instruction in Listing~\ref{list:computation_asm}, ``\verb|add A[i], A[i], 1|'', modifies the storage field $A[i]$ by adding 1 to it. It commands at least four individual resources: the datapath multiplexors (that build the correct data transfer path), the storage reading port (to read $A[i]$), the storage writing port (to write to $A[i]$), and the arithmetic unit (that performs the ``ADD-1'' operation). Computation-centric instructions are inherently sequential in the scope of the controller that issues these instructions. Instructions cannot be executed simultaneously with other instructions due to resource conflict. Even if we add more instruction issue slots like in the VLIW processors, the degree of instruction level parallelism (ILP) still cannot be improved dramatically. Fetching, decoding, and executing a computation-centric instruction is also very costly regarding power consumption because the controller is usually far away from the resources it controls; wires dominate the energy and latency cost metrics as the technology scales \cite{Kogge2013Exascale,Saraswat1982Effect}. Increasing the degree of parallelism will not reduce the cost of instructions since the total number of instruction issues will not be affected by the parallelism; each instruction has to be fetched, decoded, and executed.

\begin{minipage}{\linewidth}
\begin{lstlisting}[language=computation_asm, caption=Computation-centric pseudo assembly code, label=list:computation_asm, , xrightmargin=4.0ex]
  load i, 0
LOOP:
  compare i, 64
  branch >=, END
  add A[i], A[i], 1
  add i, i, 1
  jump LOOP
END:
  nop
\end{lstlisting}
\end{minipage}

In this paper, we introduce the Composable Instruction Set (CIS).
This instruction set uses \emph{resource-centric} instructions, which (unlike \emph{computation-centric} instructions) only configure and control a single resource. Resource-centric instructions are inherently parallel and cooperate with each other in a cycle-accurate manner. The operation emerges from the cooperative behavior of these instructions. If we compile the static C program in Listing \ref{list:c_program} to resource-centric instructions, we will get a pseudo assembly code similar to Listing \ref{list:resource_asm}. We can see that the 64-element vector addition is achieved by issuing only four instructions. It is a considerable reduction in control overhead compared to processor ISA, which needs to repeatedly issue the \verb|add| instruction and many other auxiliary instructions in a loop (see Listing~\ref{list:computation_asm}).

\begin{minipage}{\linewidth}
\begin{lstlisting}[language=resource_asm, caption=Resource-centric pseudo assembly code, xrightmargin=4.0ex, label=list:resource_asm]
Interconnect       : forever { make connection }
Storage Read Port  : repeat (i=0:64) { read A[i] }
Storage Write Port : repeat (i=0:64) { write A[i] }
Computation        : forever { perform ADD-1 }
\end{lstlisting}
\end{minipage}

Resource-centric instructions have \textbf{spatial composability}. Vector operations such as ``\emph{adding 1 to every element in a vector of 64 numbers}'' can be constructed not by repeating the same ``\verb|add|'' instruction 64 times in a loop, but by the collaboration of concurrent and independent micro-threads running on different resources.
In Listing~\ref{list:resource_asm}, four instructions compose the behavior of four micro-threads: (1) create the necessary datapath connection, (2) read the data from the vector, (3) write the computed results to the original vector, and (3) perform the ``ADD-1'' arithmetic operation. The complex vector operation emerges from the collaboration of these micro-threads, each bounded to a single resource.

As a result, using the resource-centric program (Listing~\ref{list:resource_asm}), one can perform an ``ADD-1'' operation in each cycle and fully utilize the processing unit. In contrast, using the computation-centric program (Listing~\ref{list:computation_asm}), one can only perform an ``ADD-1'' operation every five cycles, since each loop iteration is composed of five operations.

However, when mapping a complex vector operation (e.g., that contains many levels of nested loops), the resource configuration cannot be expressed with a single instruction due to the instruction's limited bitwidth.
Thus, instructions must have \textbf{temporal composablity} as well.
Temporal composability allows the construction of the description of a complex operation by combining simple instructions. For example, a complex operation ``\emph{reading elements from a vector following a 2D affine address pattern, start from address 0, inner loop repeat 3 times, outer loop repeat 5 times}'' can be constructed by the temporal collaboration of three simple resource-centric instructions: (1) \emph{read a number}, (2) \emph{repeat 3 times with step=1}, and (3) \emph{repeat 5 times with step=1}. In contrast, conventional processor instructions are not temporally composable because each instruction exists independently. In our example, instruction (2) or (3) (i.e., ``\emph{repeat}'' operations) cannot exist independently from instruction (1) (i.e., ``\emph{read a number}'' operation). Hence, temporally-composable resource-centric instructions must collaborate.

We can see from the motivational example above that the composable instruction set (CIS), in both spatial and temporal sense, can naturally express different loop structures in a distributed and concurrent manner.
Hence, CIS is ideal for accelerating streaming applications that are typically dominated by static loops.
Therefore, in this paper, we conduct a detailed analysis of CIS, focusing solely on instruction set design. 

In summary, this work proposes the following contributions:
\begin{itemize}
    \item We introduce the composable instruction set (CIS), a novel resource-centruc programming approach, well-suited for data streaming applications, and provide a toy example for demonstration purposes (Section~\ref{sec:3}).
    \item We compare the resource-centric and composable approach of CIS with current mainstream state-of-the-art ISAs for typical streaming applications (Section~\ref{sec:4}).
\end{itemize}


%% file: contents/sec2.tex
\section{State-of-the-Art}\label{sec:2}
Processor instruction set architectures (ISAs) can be categorized into two families: Complex Instruction Set Computer (CISC) and Reduced Instruction Set Computer (RISC). The community has well acknowledged that the shifting from the CISC (e.g., x86~\cite{dandamudi2013introduction}) to RISC (e.g., MIPS~\cite{britton2002mips}, ARM~\cite{knaggs2004arm}, RISC-V~\cite{waterman2014risc}) generally improves the efficiency of processors because the RISC ISAs greatly reduce the complexity of controller design by reducing the number of instruction types \cite{patterson1980the}.
However, processor ISAs focus on computation instead of configuration of each hardware resource.
As shown in the motivational example (see Section~\ref{sec:1}), these computation-centric instructions are inherently sequential and cannot implement loop structure efficiently.
The centralized controller also increases the cost of instruction issues because the control points in the datapath are far away from the centralized controller.

The micro-programmable instructions~\cite{rauscher1980microprogramming} look similar to temporally composable instructions, but they are, in fact, very different. The controller that supports microprogrammable instructions translates macro instructions into sequences of micro instructions. Each micro instruction is an independent instruction. They don't cooperate with other micro instructions to create complex control structures (e.g., nested loop structures).

Architectures like Graphic Processing Unit (GPU) and RISC-V based vector extensions (e.g., Ara~\cite{Perotti2024ara2}) use single instruction multiple data (SIMD) instructions~\cite{franchetti2005efficient}.
Their construction is very similar to standard processor instructions, except that a single instruction will perform a vector operation distributed spatially.
However, SIMD is limited as it can only tackle simple shallow loops and has a fixed vector size.

Coarse-grain reconfigurable architectures (CGRAs) use NoC-based grid-like architecture to connect many processing elements (PEs) to increase the degree of parallelism.
However, CGRA compilers (e.g. \cite{Regheb2024CGRA-ME,Tirelli2023SAT-MapIt}) only support mapping the innermost loop structure of a program on target CGRA platforms by using software pipelining technique.
Many PEs are used as address generation and routing devices instead of effective arithmetic computation units.

Very long instruction word (VLIW) architectures leverage parallelism to accelerate the program. However, as explained in \cite{hennessy2011computer}, high instruction level parallelism (ILP) is difficult to achieve in VLIW. It is very difficult to develop efficient VLIW compilers to reduce the number of NOPs and increase device utilization.

Spatially composable ISAs have been proposed in the literature.
The persistent and fully cooperative instructions proposed in~\cite{yang2021scheduling,yang2022reducing} are resource-centric.
However, the authors did not address the need for temporal composability to accommodate complex operations.
Instead, the authors used complex CISC-like long instructions, significantly complicating the controller decoding logic. In~\cite{catthoor2010ultra}, the authors proposed a spatially composable ISA by introducing the concept of loop buffers for each resource.
The programmable loop buffer is similar to a processor controller, which makes it very generic and flexible.
However, it did not address complex operation configuration problems. It also needs extra instructions for secondary tasks like address computation inside the loop structures.

Some instruction sets with data-driven control logic have also been proposed in the literature, such as~\cite{parashar2013triggered}.
However, these instruction sets suffer similar limitations as processor ISAs as they have high control overhead. This makes them unsuitable for static data streaming application acceleration.

In constrast, the spatially composable ISA approach simplifies the control logic compared to processors and introduces concurrent and collaborative micro threads to avoid control hazards as much as possible. It aims to accelerate not only the innermost loop but many layers of nested loops. With dedicated resources for address generation and path routing, CIS can utilize PEs much more efficiently compared to other architectures. The temporal composability also helps to simplify the controller design compared to other ISAs that is only spatially composable.

%% file: contents/sec3.tex
\section{Composable Instruction-Set Architecture}\label{sec:3}
In this section, we describe the characteristics of a composable instruction set (CIS) architecture. To lay the foundation, we first explain the hardware architecture concept, followed by the concept of spatial and temporal composability in the context of ISA design. Then, we detail a toy example to support our description of the CIS concept. Finally, we discuss the hardware and compiler design implications of CIS. This paper focuses on the instruction set design aspects of CIS. Hence, the hardware architecture design and the compiler design aspects will only be described from a high-level of abstraction, only detailing the implementation details required to understand CIS.

\subsection{Hardware Architecture Template}

\begin{figure}[!ht]
    \centering
    \includegraphics{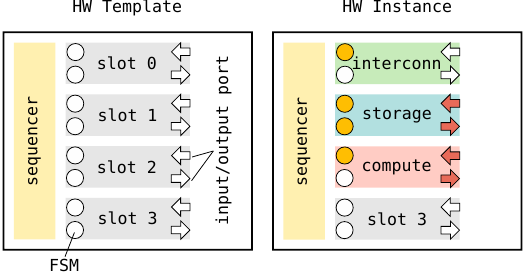}
    \caption{The hardware architecture is a template that consists of a sequencer and multiple resource slots. The hardware architecture instance will be used to demonstrate later examples. Colored slots are occupied by resources, but gray slots are unused. Resources can occupy multiple continuous slots, hence accessing multiple ports.}
    \label{fig:hw}
\end{figure}

Here, we describe the conceptual hardware architecture on which CIS will run.
As shown in Fig.~\ref{fig:hw}, the hardware architecture template is composed of a \textbf{sequencer} (i.e., its controller) and several resource slots.
Each resource slot has two local final-state machines (FSMs) and two ports: one for input and one for output.
Each resource, depending on its implementation, can use part or all of its FSMs and ports.
The minimal hardware architecture instance that we use as a running example includes three resources: a \textbf{computation unit (C)}, an \textbf{interconnection unit (I)}, and a \textbf{storage units (S)}.

The hardware template shown in Fig.~\ref{fig:hw} naturally forms a single computation and/or storage tile. We can construct a CGRA-style \emph{fabric} by connecting those tiles to a 2D grid. For the toy example (see Section~\ref{sec:toy}), we focus on a single tile. However, in our experiments (see Section~\ref{sec:4}), we also compare multi-tile instances of this template to the state-of-the-art.

\subsection{Spatial Composability}
Spatial composability refers to the ability to construct a complex operation from many simple atomic operations distributed on different hardware resources.
These simple atomic operations are self-contained and autonomous.

An instruction set with spatial composability is necessarily resource-centric.
This means that each instruction from the instruction set targets a specific hardware resource.
For example, while one instruction could configure the interconnect to build the path for the operands (and return values) of a function to be mapped, another instruction could perform a read operation to deliver one of the operands, and a third different instruction could configure the arithmetic logic unit (ALU) to perform the desired transformation function (e.g., addition, multiplication).
Thus, the complete operation is composed from many different smaller operations (e.g., building data transfer paths, reading operands, performing arithmetic functions).
Each small task is carried out by a specific instruction that interacts only with a particular set of hardware resources. For a more concrete example, refer to the example in Section~\ref{sec:toy}.

\subsection{Temporal Composability}
Temporal composability refers to the ability to construct a complex operation from one or more atomic basic operations and some well-defined temporal transformation operators.
The basic operations are standalone and can configure certain resources to a specific state. They are usually single-cycle operations.
The temporal transformation operators can transform the timing properties of the basic operation and must be used with one or more basic operations.

The temporal property of an operation can usually be represented by a finite-state machine (FSM).
A basic operation can be treated as an event that forms a state in the FSM, and the temporal transformation operators create the transition edges in the FSM.
Arbitrary FSMs are challenging to construct using a minimal set of transition patterns.
Since we specifically target streaming applications in this paper, we only need two temporal transformation operators to implement most FSMs for streaming applications: a ``REPETITION'' operator and a ``TRANSITION'' operator.

The ``REPETITION'' (\(\mathbf{R}\)) operator accepts only one instruction block as argument.
It repeats its inner instruction block multiple times.
This operator corresponds to the FSM that implements the FOR-LOOP structure.
A single $\mathbf{R}$ operator mimics a single layer of FOR-LOOP.
\(\mathbf{R}\) operators can be nested to implement a multi-level FOR-LOOP structure.

The ``TRANSITION'' (\(\mathbf{T}\)) operator accepts two instruction blocks as arguments.
After a specific delay, it forces the transition from the first inner block to the second inner block.
This operator corresponds to the FSM that implements a two-number counter.
\(\mathbf{R}\) operators can be stacked to implement an N-number counter FSM.

When a streaming application requires more complex control patterns (e.g., WHILE-LOOP, IF-THEN-ELSE) that cannot be directly decomposed to the combination of $\mathbf{R}$ and $\mathbf{T}$ operators.
Hence, we use traditional control instructions like \verb|COMPARE|, \verb|BRANCH|, or \verb|JUMP| instructions to implement the outer control flow.
Thus, the simpler inner control flow can be expressed by the combination of $\mathbf{R}$ and $\mathbf{T}$ operators.
Note that the inner control flow is not theoretically limited in the number of loop layers that it can represent.
Hence, unlike GPUs and CGRAs (see Section~\ref{sec:2}), CIS is not limited to accelerating only the innermost loop of a program.
In practice, we observe that CIS typically accelerates up to 4 layers of nested loop.

\subsection{Toy Example}\label{sec:toy}
Here, we define a toy example of a CIS instance.
This example consists of seven instructions divided into three categories: resource, transform, and control (as shown in Table~\ref{tab:isa}).
Note that this toy example is only for demonstration purposes.
The actual CIS instances used in Section~\ref{sec:4} for experiments implement more complex and scaled-up versions of this example, but are based on the same architectural template.

\begin{table}[!ht]
    \centering
    \caption{Example of a toy CIS consisting of only seven instructions.}
    \label{tab:isa}
    \begin{tabular}{|l|l|l|}
    \hline
    Type & Name & Format and Description \\
    \hline
    \multirow{6}{*}{Resource} & \verb|@C| & \verb|@C [slot:FSM] [option] [function]| \\
    & & Configure a \textbf{computation} resource.\\ \cline{2-3}
    & \verb|@I| & \verb|@I [slot:FSM] [option] [path]|\\
    & & Create an \textbf{interconnection} between ports.\\ \cline{2-3}
    & \verb|@S| & \verb|@S [slot:FSM] [address]| \\
    & & Create a read/write for a \textbf{storage} unit.\\
    \hline
    \multirow{4}{*}{Transform} & \verb|@R| & \verb|@R [slot:FSM] [iter] [step] [delay]| \\
    & & Implementation of the $\mathbf{R}$ operator.\\ \cline{2-3}
    & \verb|@T| & \verb|@T [slot:FSM] [delay]| \\
    & & Implementation of the $\mathbf{T}$ operator.\\
    \hline
    \multirow{4}{*}{Control} & \verb|@W| & \verb|@W [delay]| \\
    & & The controller \textbf{waits} for specific cycles.\\ \cline{2-3}
    & \verb|@A| & \verb|@A [slot:FSM list]| \\
    & & \textbf{Activate} FSMs in the list.\\
    \hline
    \end{tabular}
\end{table}

\begin{figure}[!ht]
    \centering
    \includegraphics{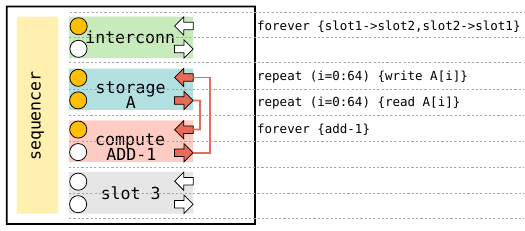}
    \caption{The decomposition of program to operations}
    \label{fig:option}
\end{figure}

We use these instructions to map the simple vector addition previously shown in Listing~\ref{list:c_program}.
We must configure four FSMs to perform the vector addition: interconnection, storage read, storage write, and computation.
The main FOR-LOOP is naturally decomposed into four concurrent operations shown in Fig.~\ref{fig:option}.
While some operations have a loop structure, some do not.
For example, the FSM in slot 0 (i.e., interconnect) will build a connection from slot 1 (i.e., storage) to slot 2 (i.e., compute unit) and from slot 2 to slot 1.
This operation does not require repetition because once a connection is established, it will remain there until it is reconfigured.
In contrast, the FSM in slot 1 must repeat its event 64 times because it must write the result of each computation back to the storage unit.
Note that even if many local FSMs are running in parallel, the hardware architecture is not similar to VLIW or superscalar architecture whose controllers are multi-issue.
The controller (i.e., sequencer) in our architecture template is single-issue, making its implementation relatively easy and economic.

To further detail this toy example, we provide the complete assembly program that implements the vector addition as shown in Listing~\ref{list:assembly}.

\begin{minipage}{\linewidth}
\begin{lstlisting}[caption=CIS assembly code, label=list:assembly, language=asm, xrightmargin=4.0ex]
# Configure the interconnection path; both instructions configure option0 because both paths should exist simultaneously as configuration option 0.
@I slot0:FSM0 option0 slot1->slot2
@I slot0:FSM0 option0 slot2->slot1
# Set the computation unit to perform +1
@C slot2:FSM0 option0 ADD-1
# Configure the output port for the storage unit to generate addresses from 0 to 63 
@S slot1:FSM1 address=0
@R slot1:FSM1 iter=64 step=1 delay=0
# Configure the input port for the storage unit to generate addresses from 0 to 63
@S slot1:FSM0 address=0
@R slot1:FSM0 iter=64 step=1 delay=0
# Activate the interconnection and computation unit
@A [slot0:FSM0, slot2:FSM0]
# Activate the output port of the storage unit
@A [slot1:FSM1]
# Wait for a cycle to allow the result data to be computed
@W delay=1
# Activate the input port of the storage unit
@A [slot1:FSM0]
# Wait until all operations are finished
@W delay=63
\end{lstlisting}
\end{minipage}

\subsection{Impact on architecture and compiler design}
CIS can impact its compiler design, mainly negatively.
Since the hardware that supports CIS is massively parallel due to all those distributed local controllers on each resource, the compiler must schedule the instructions in a cycle-accurate manner to guarantee the correct instruction cooperation and produce the right computation results.
This imposes high requirements on the compiler instruction scheduler.
The extendability requirement of CIS-based solutions further complicate compiler design, since the compiler or even the whole software eco-system has to be designed in a modular fashion so that new resources can be added as extensions.

The CIS architecture can also positively and negatively impact hardware architecture design. On the one hand, CIS requires the hardware platform to be designed as a template to allow flexible resource expansion.
This modularity requirement might complicate the hardware design process as some specific functionalities may require more CIS instructions to implement compared to conventional ISAs since operations are decomposed to fine-grained atomic instructions.
For example, the $\mathbf{R}$ operator that represents the same shared loop structure has to be repeated on each resource since their local controllers are not shared.
On the other hand, CIS is naturally designed for heterogeneous computing.
Each resource slot can host any resource necessary for the specific application.
Each resource only accepts a small subset of the ISA (typically 2-3 instruction types).
The reduced instruction type drastically simplifies the hardware decoding logic.
Placing the control FSM near the resources reduces the length of control wires, thus greatly reducing power consumption~\cite{Kogge2013Exascale,Saraswat1982Effect}.

%% file: contents/sec4.tex
\section{Experiment}\label{sec:4}
In this section, we present experimental results comparing the CIS approach to other approaches from the state-of-the-art.
As described in the beginning of Section~\ref{sec:3}, this paper focuses on the instruction set design.
Thus, the goal of our experiments is to demonstrate the efficiency of the CIS concept without diving into specific details of hardware architecture.
In the following experiments, we use a
CGRA-like architecture whose resources can be configured at compile-time, that implements CIS.
We call this architecture Dynamic Reconfigurable Resource Array (DRRA).
While the CIS used to produce the following results is also significantly more complex than the toy example presented in Section~\ref{sec:3}, it implements the same CIS concepts.

\begin{table}[!ht]
    \centering
    \caption{Hardware Architectures for Comparison}
    \label{tab:machines}
    \begin{tabular}{|l|l|l|l|}
         \hline
         Architecture Feature & Representative Machine & Resources & Literature \\ \hline\hline
         CIS  & DRRA & 1 PE / 8 PEs & This work \\
         Micro-processor  & RISC-V & 1 PE & \cite{picorv32} \\
         SIMD & ARA-2 & 8 PEs & \cite{Perotti2024ara2}\\
         CGRA & OpenEdge CGRA& 4 PEs / 8 PEs & \cite{Alvarez2023An,Tirelli2023SAT-MapIt}\\
         VLIW & TI C7000& 8 PEs & \cite{tic7000}\\
         \hline
    \end{tabular}
\end{table}

We choose a set of representative hardware architectures to compare with the DRRA architecture.
To test the benefits of CIS compared to various state-of-the-art styles of instruction set, we map different data streaming applications to the architectures listed in Table~\ref{tab:machines}.
The application set for the comparison includes: dot product (DOT), matrix-vector multiplication (MVM), matrix-matrix multiplication (MMM), 1D convolution (1DCONV), and 2D convolution (2DCONV).
For each application, two instances of different sizes are implemented.

We do not directly compare latency ($L$) since the latency number is affected by many factors (e.g., the algorithm complexity and the degree of parallelism), making it less straightforward to show the efficiency of the instruction set.
We instead compute the effective Processing Element (PE) utilization ($\eta_{eff}$) by Eq~\ref{eq:eta}.

\begin{equation}
    \eta_{eff} = \frac{N_{PE} \times L}{\Omega_{eff}}
    \label{eq:eta}
\end{equation}

Where, $\Omega_{eff}$ is the \emph{effective operation count} of a specific algorithm, which means the number of \emph{core} operations.
For example, the effective operations of a 32-element dot-product algorithm are 32 multiplication operations and 32 addition operations (i.e., $\Omega_{eff}(DOT\_32) = 64$).
The $N_{PE}$ is the \emph{number of PEs} that can be used to execute these \emph{effective} operations.
We assume that each effective operation takes one cycle to execute on a single PE.
From Eq.~\ref{eq:eta}, we can see that the $\eta_{eff}$ represents the efficiency of the instruction set design.
It largely removes the influence from the degree of parallelism ($N_{PE}$) and the algorithm complexity ($\Omega_{eff}$) from the measured latency ($L$).
A good instruction set should automatically resolve resource dependencies and avoid starvation of PEs, thus improving $\eta_{eff}$.
In subsequent figures, the applications are sorted by their theoretical number of core operations, from lowest to highest.

\subsection{Compare with Micro-processor}

\begin{figure}[!ht]
    \centering
    \includegraphics[width=\textwidth]{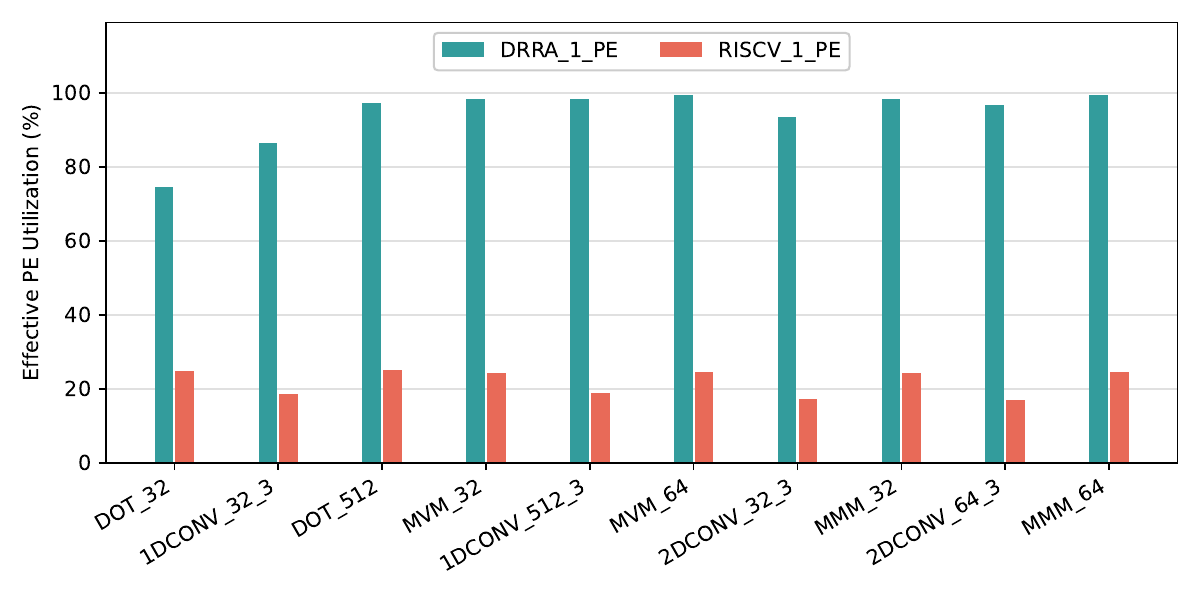}
    \caption{The effective PE utilization comparison: DRRA vs RISC-V}
    \label{fig:drra_vs_riscv}
\end{figure}

The comparison result between DRRA and RISC-V micro-processor is shown in Fig.~\ref{fig:drra_vs_riscv}. It's clear that CIS-based DRRA has much higher $\eta_{eff}$ compared to RISC-V. The CIS instructions divide the mapped algorithm into concurrent micro-threads thus eliminate the control hazards caused by limited instruction issue slots. The PE thus can be utilized much more since all the resources are configured to work autonomously to make the data delivery to the PE much more efficient. Note that, the $\eta_{eff}$ of DRRA almost reaches the theoretical maximum as long as the mapped algorithms are not trivially simple. The DRRA result is excellent considering that DRRA is a domain-specific programmable architecture, not a hand-optimized ASIC-style machine.

\subsection{Compare with Parallel Architectures}

\begin{figure}[!ht]
    \centering
    \includegraphics[width=\textwidth]{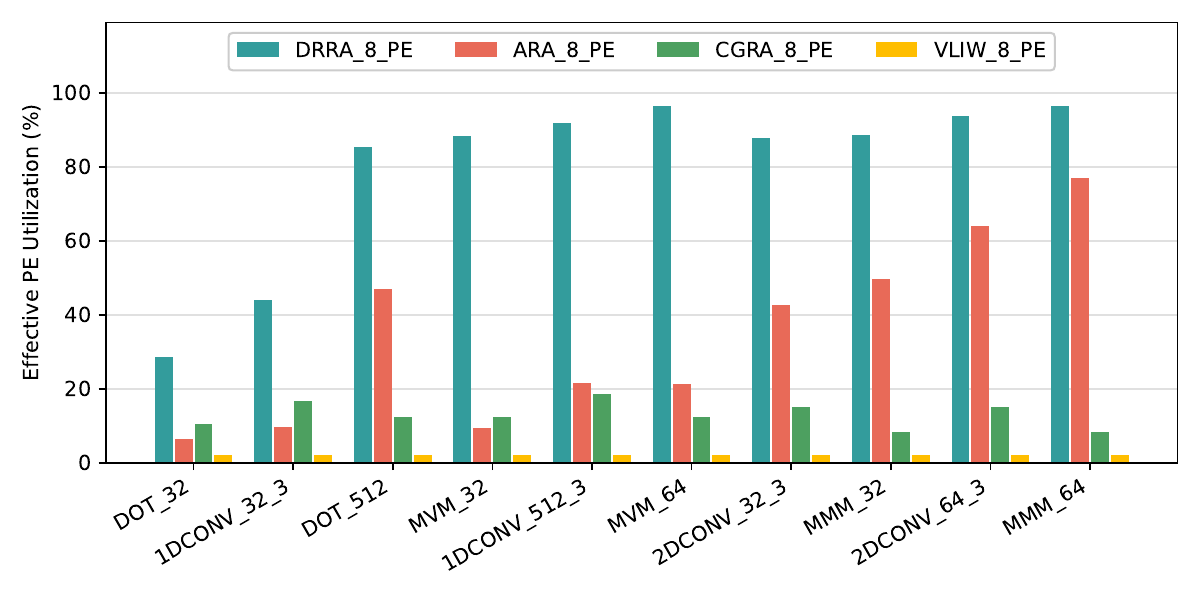}
    \caption{The effective PE utilization comparison: DRRA vs ARA-2, OpenEdge CGRA and TI C7000 VLIW}
    \label{fig:drra_vs_ara_cgra_vliw}
\end{figure}

We also compare the DRRA with parallel architectures like SIMD (ARA-2), CGRA (OpenEdge CGRA) and VLIW (TI C7000). All architectures are equipped with 8 PEs. The comparison result is shown in Fig.~\ref{fig:drra_vs_ara_cgra_vliw}. From the graph we can easily observe that DRRA is outperform all of those architectures. The SIMD performs relatively good when the algorithm is complex. The effective PE utilization of CGRA and VLIW are quite low and it does not improve even when the algorithm is complex.

\subsection{The Effect of Parallelism}

\begin{figure}[!ht]
    \centering
    \includegraphics[width=\textwidth]{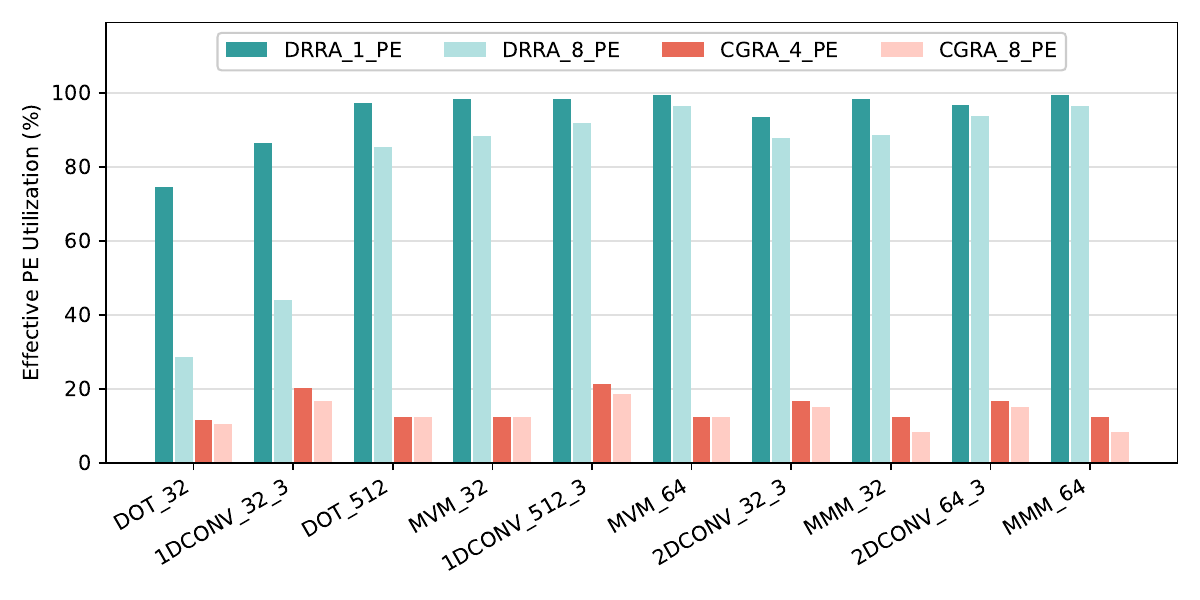}
    \caption{The effect of parallelism on both DRRA and CGRA}
    \label{fig:parallelism}
\end{figure}

As shown in Fig~\ref{fig:parallelism}, we can clearly see that \emph{increasing the degree of parallelism by adding more processing elements (PEs) will not increase the effective PE utilization ($\eta_{eff}$). On the contrary, it usually lower the utilization.} The utilization drops the most usually when the algorithms are simple, i.e. $\Omega_{eff}$ are small. The reason of utilization dropping is that keeping more PEs busy all the time is much more challenging than just keeping a single PE busy. Therefore, parallel architectures like SIMD, CGRA, and VLIW with a higher degree of parallelism will not magically eliminate the inefficiency inherent in the computation-centric ISA.

Readers might noticed that the $\eta_{eff}$ of ARA-2 is higher than that of RISC-V, which seems inconsistent with our conclusion regarding the effect of parallelism, considering that ARA-2 uses RISC-V as its host CPU.
However, this is simply because when mapping algorithm on ARA-2, the algorithm is hand-optimized using techniques like loop unrolling, which reduces overhead per effective operation.

We also want to point out that even though our comparison focuses on the effective PE utilization, the conclusion also applies to applications that are traditionally considered as ``memory-bound''.
Those applications are memory-bound only because the conventional architecture cannot provide enough memory bandwidth, leading to PE starvation.
A memory-bound application can become computation-bound once the memory bottleneck is removed.
Our hardware architecture template could improve on this front as well.
It can be extended to a tiled architecture with multiple I/O ports.
SRAM blocks can also be treated as regular resources embedded in the tiled architecture and thus can be used as scratchpad memory wherever needed.

%% file: contents/sec5.tex
\section{Conclusion and Future Works}\label{sec:5}
This paper introduces a novel and highly efficient Composable Instruction Set (CIS) architecture designed for data streaming applications.
The proposed CIS significantly improves effective PE utilization, approaching its theoretical maximum.
Future research will quantitatively evaluate the impact of the CIS on hardware design and benchmark its performance against existing hardware platforms, specifically focusing on power consumption.
Furthermore, we aim to develop robust compilation support, particularly instruction scheduling algorithms, to enhance the usability and accessibility of the CIS-based platform.